\documentclass[fleqn,usenatbib]{mnras}

\usepackage{newtxtext,newtxmath}
\usepackage[T1]{fontenc}
\usepackage{ae,aecompl}

\usepackage{natbib}

\usepackage{graphicx}	% Including figure files
\usepackage{amsmath}
\usepackage{hyperref}
\usepackage{listings}
\usepackage{color}
\usepackage{float}
\usepackage{xspace} 
\usepackage{longtable}

\newcommand{\kepler}{\textsl{Kepler}\xspace}

\title[Stellar Variability in Exoplanet Transits with PLATO]{The Stellar Variability Noise Floor for Transiting Exoplanet Photometry with PLATO}

\author[Morris et al.]{Brett M. Morris,$^{1}$\thanks{E-mail: morrisbrettm@gmail.com}
Monica G. Bobra,$^{2}$ Eric Agol$^{1}$, Yu Jin Lee,$^{3}$ Suzanne L. Hawley$^{1}$\\
$^{1}$Astronomy Department, University of Washington, Seattle, WA 98195, USA\\
$^{2}$W. W. Hansen Experimental Physics Laboratory, Stanford University, Stanford, CA 94305, USA\\
$^{3}$Stanford University, Stanford, CA 94305, USA\\
}

\date{Accepted XXX. Received YYY; in original form ZZZ}

\pubyear{2018}

\begin{document}
\label{firstpage}
\pagerange{\pageref{firstpage}--\pageref{lastpage}}
\maketitle

\begin{abstract}
One of the main science motivations for the ESA PLAnetary Transit and Oscillations (PLATO) mission is to measure exoplanet transit radii with 3\% precision. In addition to flares and starspots, stellar oscillations and granulation will enforce fundamental noise floors for transiting exoplanet radius measurements. We simulate light curves of Earth-sized exoplanets transiting continuum intensity images of the Sun taken by the Helioseismic and Magnetic Imager (HMI) instrument aboard the Solar Dynamics Observatory (SDO) to investigate the uncertainties introduced on the exoplanet radius measurements by stellar granulation and oscillations. After modeling the solar variability with a Gaussian process, we find that the amplitude of solar oscillations and granulation is of order 100 ppm -- similar to the depth of an Earth transit -- and introduces a fractional uncertainty on the depth of transit of 0.73\%
assuming four transits are observed over the mission duration.  However, when we translate the depth measurement into a radius measurement of the planet, we find a much larger radius uncertainty of 3.6\%.  This is due to a degeneracy between the transit radius ratio, the limb-darkening, and the impact parameter caused by the inability to constrain the transit impact parameter in the presence of stellar variability. We find that surface brightness inhomogeneity due to photospheric granulation contributes a lower limit of only 2 ppm to the photometry in-transit. The radius uncertainty due to granulation and oscillations, combined with the degeneracy with the transit impact parameter, accounts for a significant fraction of the error budget of the PLATO mission, before detector or observational noise is introduced to the light curve. If it is possible to constrain the impact parameter or to obtain follow-up observations at longer wavelengths where limb-darkening is less significant, this may enable higher precision radius measurements.
\end{abstract}


\section{Introduction}

The detection of Earth-like planets around Sun-like stars is a long-standing goal of astronomy.  The {\it Kepler} spacecraft set out to discover Earth-Sun analogs, and its most promising discovery was a transiting planet on an orbit similar to Earth around a star similar to our Sun, Kepler-452b \citep{Jenkins2015}, although the statistical significance of this detection has been debated \citep{Mullally2018, Burke2019}.

In addition to the orbital period, the radius-ratio of the planet to the star may be inferred from the transit depth after accounting for limb-darkening.  When coupled to estimates of the radius of the star, this yields an estimate of the planet radius, which is an indicator of the possible composition of the planet, as revealed by transiting planets at shorter orbital periods \citep{Rogers2015,Fulton2017}.  The planet Kepler-452 b has a radius about 60\% larger than Earth, which may imply a different composition \citep{Rogers2015}. Other promising candidates exist with shorter periods and/or
larger radii than Earth, as well as host stars which are different from our Sun;  we have not detected a true Earth-Sun analog.

The transit technique still holds promise for the detection of closer Earth-Sun analogs. The ESA PLAnetary Transits and Oscillations (PLATO) mission aims to discover transiting exoplanets among 15,000 dwarf and sub giant stars with $V<11$ with 45 second cadence photometry collected over 4.5 years \citep{Rauer2014}. One of the main scientific motivations for PLATO is to measure Earth-like exoplanet transit radii with 3\% precision. With this precision, we may be able to answer such compelling questions as: do planetary radii evolve with stellar age; can we accurately measure the dearth of planets near the radius gap at $2R_\oplus$ \citep{Fulton2017}?  To achieve this ambitious goal, the host stars likely must have little stellar activity in the form of starspots or flares, and the stellar variability must be accounted for in the transit analysis.  

Here we examine the noise floor on measuring the planet-star radius ratio which is caused by stellar inhomogeneity and variability coupled with stellar limb-darkening.  Two forms of variability during transits may affect transit measurements:  variability caused by non-uniformity in the surface brightness of the star which is being occulted by the planet, and variability due to temporal variations in the total flux emitted by the unocculted star.  Limb-darkening affects the depth of the transit depending on how close the path of the transit passes with respect to the center of the star.  These three effects limit how well we can measure the planet-star radius ratio.

In principle, space-based photometry of transiting exoplanets contains information about the host star as well as the exoplanet. Light curves of transiting exoplanets have been used to map stellar surface brightness variations within the transit chord, occasionally revealing maps of starspots \citep[see e.g.][]{Pont2007, Pont2013, Desert2011, Sanchis-Ojeda2011, Bonomo2012, McCullough2014,Oshagh2014, Davenport2015thesis, Morris2017a, Morris2018b}.  As a result, one must be careful to account for stellar surface inhomogeneities when measuring the radius of an exoplanet \citep{Morris2018}.  

The Sun and Sun-like stars exhibit photospheric granulation, a small-amplitude surface inhomogeneity observable in transit photometry \citep{Chiavassa2017}. Stellar granulation is the pattern of hot, bright upwelling convective plasma surrounded by cooler and dimmer down-flowing plasma which tiles the photosphere of a Sun-like star. Granulation occurs on a variety of size scales. The most prominent in typical continuum images of the Sun are granules with radii of $\sim 0.5$ megameter (Mm), or roughly $0.08 R_\oplus$; less obvious is the structure of supergranules, which span $\sim 16$ Mm in radius, or roughly $2.5R_\oplus$ \citep{Rast2003, Chiavassa2017}. 

The degree to which granules and supergranules affect transit light curves depends on their size scales and intensity contrast. Several granules are occulted by an Earth-sized planet -- 6.37 Mm in radius -- at any instant during an exoplanet transit event. However, only a portion of a supergranule is occulted during an exoplanet transit. Furthermore, it is easy to observe granulation in continuum intensity data (see Figure~\ref{fig:hmi}), whereas the relative contrast between the bright, upwelling plasma and dark, down-flowing plasma in supergranular cells is much lower than for granular cells \citep{Nordlund2009}. Nevertheless, transits of Earth-sized exoplanets could in principle reveal the pattern of stellar supergranulation in the residuals of extremely high precision transit photometry.  However, as we show in this paper, this will be swamped by temporal stellar variability.

The Sun and Sun-like stars also exhibit acoustic pressure mode, or $p$-mode, oscillations driven by convection near the solar surface \citep{Christensen-Dalsgaard2002}. Spatially-resolved Doppler velocity and photometric measurements of the Sun reveal $p$-mode oscillations with a maximum power near period $P=5.39 \pm 0.05$ minutes \citep{Fröhlich1997, Huber2011}, often referred to as the solar ``five-minute oscillations.'' These oscillations are omnipresent in solar and stellar photometry, injecting a continuous source of correlated astrophysical noise into transiting exoplanet light curves, which imposes a noise floor on transit depths -- and therefore planet radii -- measured from photometry. 

There is modest literature on simulating space-based transit photometry along with realistic stellar variability, see for example \citet{Aigrain2004a,Aigrain2004b,Carpano2008,Hippke2015}.  In this paper, we seek to build on those works by studying the specific case of an Earth-sized planet transiting using actual observations of the Sun, and by examining the fundamental noise limits on the measurement of the radii of planets from transits by Earth-Sun analogs.

The continuum intensity image time series from the Helioseismic and Magnetic Imager (HMI; \citealt{schou12}) aboard the Solar Dynamics Observatory (SDO) provides us with a useful dataset for simulating transits of exoplanets on observations of the Sun. We discuss the HMI instrument in Section~\ref{sec:hmi}.  There are two stellar contributions to the uncertainty in a transit measurement:  spatial variations in the surface brightness of the star which is occulted by the planet (\S \ref{sec:sdohmi}) and temporal variations in the brightness of the unocculted star (\S \ref{sec:photometry}).
In this work, we artificially superimpose an exoplanet on HMI continuum intensity images to create simulated transits of the real Sun. We then analyze these transits, which contain the signatures of solar granulation and $p$-mode oscillations, and, in Sections~\ref{sec:sdohmi} and \ref{sec:photometry}, we quantify the impact of each component on the exoplanet radius uncertainty. We discuss the results in Section~\ref{sec:discussion} and summarize in Section~\ref{sec:conclusion}.

\section{The HMI Instrument} 
\label{sec:hmi}
\begin{figure}
    \centering
    \includegraphics[scale=0.3]{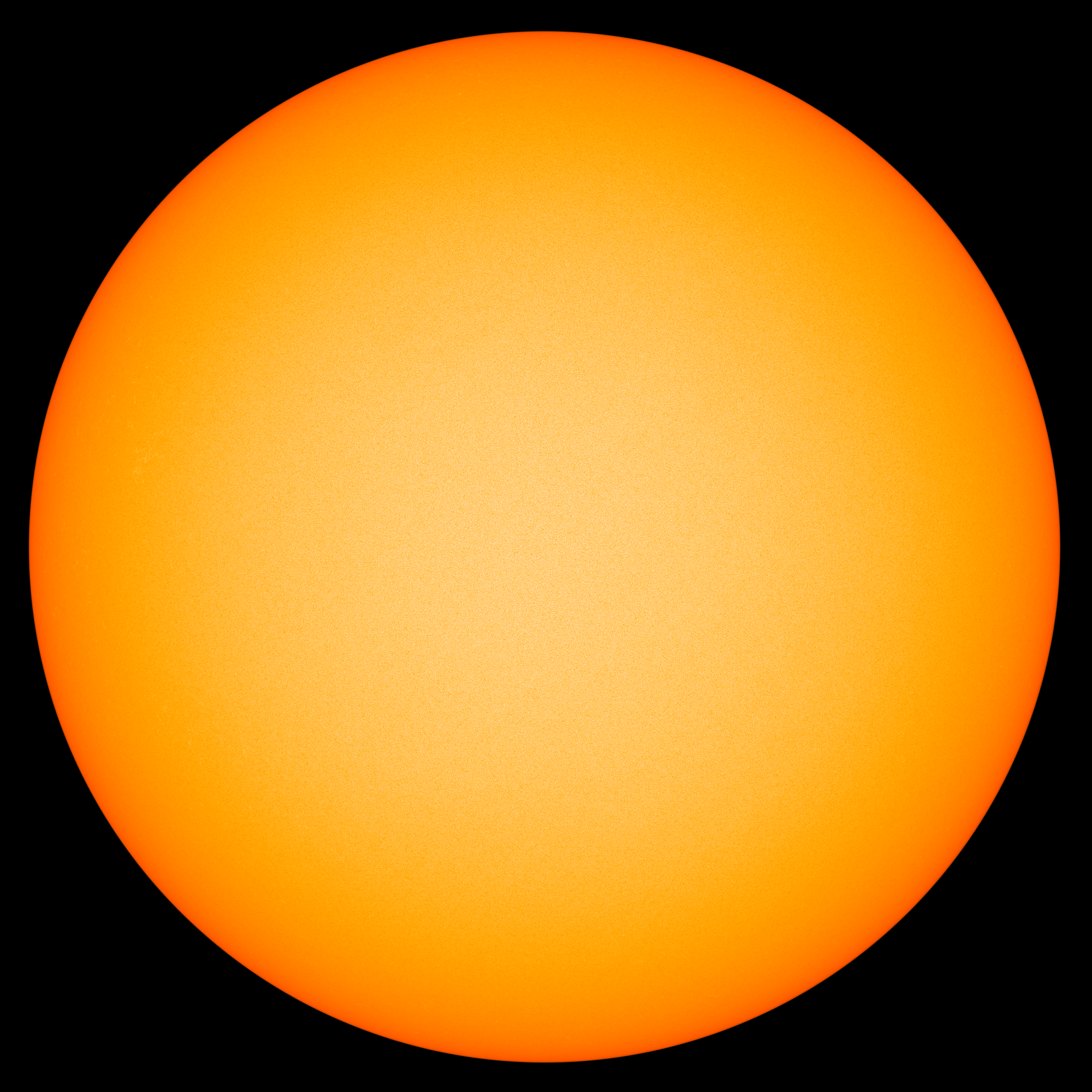}
    \caption{Deconvolved SDO/HMI continuum intensity image of the quiet Sun on 21 January 2018 at 19:34 TAI with the color table scaled from 0 to 65000 DN/s.}
    \label{fig:hmi}
\end{figure}

The HMI instrument aboard SDO takes a series of images in 6 wavelengths, centered on the Fe I spectral line at 6173 \AA, and 4 polarizations, which are then processed together to derive physical observables \citep{schou12}. One observable is the continuum intensity (see Figure~\ref{fig:hmi}), which is derived by reconstructing the solar line from the Doppler-shift, line-width, and line-depth estimates following Equation 14 of \citet{couvidat16}, who show that the continuum intensity derived in this manner differs from the true continuum intensity by less than 1\%.

Photon noise accounts for approximately 0.01\% of these continuum intensity measurements \citep{couvidat16}; similarly, systematic variations in the flat field on the order of 100 ppm add an additional source of noise to the data. These numbers exceed the performance specifications for the instrument.

As with any telescope, optical limitations also affect the image quality. Ideally, the light from a singular point on the Sun corresponds to a singular point on an image of the Sun. In practice, diffraction and imperfect optics cause light from a singular point on the Sun to spread over many points on an image of the Sun \citep{couvidat16}. This spread is characterized by a Point Spread Function (PSF). To mitigate the effects of diffraction and imperfect optics, we deconvolve HMI continuum intensity images of the Sun with the PSF (modeled as an Airy function; for details, see \citealt{wachter12}). This process increases the granular contrast by a factor of two (see Figure 27 of \citealp{couvidat16}), and ensures that optical artifacts play a small role in the following simulated transits. 

We also remove the diurnal variation in the amplitude of the HMI continuum intensity measurements, which appear due to the orbital velocity of the SDO spacecraft (see Section 2.10 of \citealp{couvidat16}), that would otherwise dominate over the small-scale perturbations we are trying to measure in this work. There is no consensus on the optimal correction method to reduce the presence and effects of the 24-hour orbital period in the SDO data. Several correction methods are outlined in Section 3.2 of \citealp{couvidat16} as well as \citealp{Schuck16}. We discuss our methods for removing this signal in Sections~\ref{sec:sdohmi} and \ref{sec:photometry}. The removal of the variations on diurnal timescales also corrects for longer term variability when comparing observations taken across different years such as degredation of the telescope's window opacity.

All the HMI data are publicly available at \url{http://jsoc.stanford.edu}.

\begin{figure}
    \centering
    \includegraphics[scale=2.475]{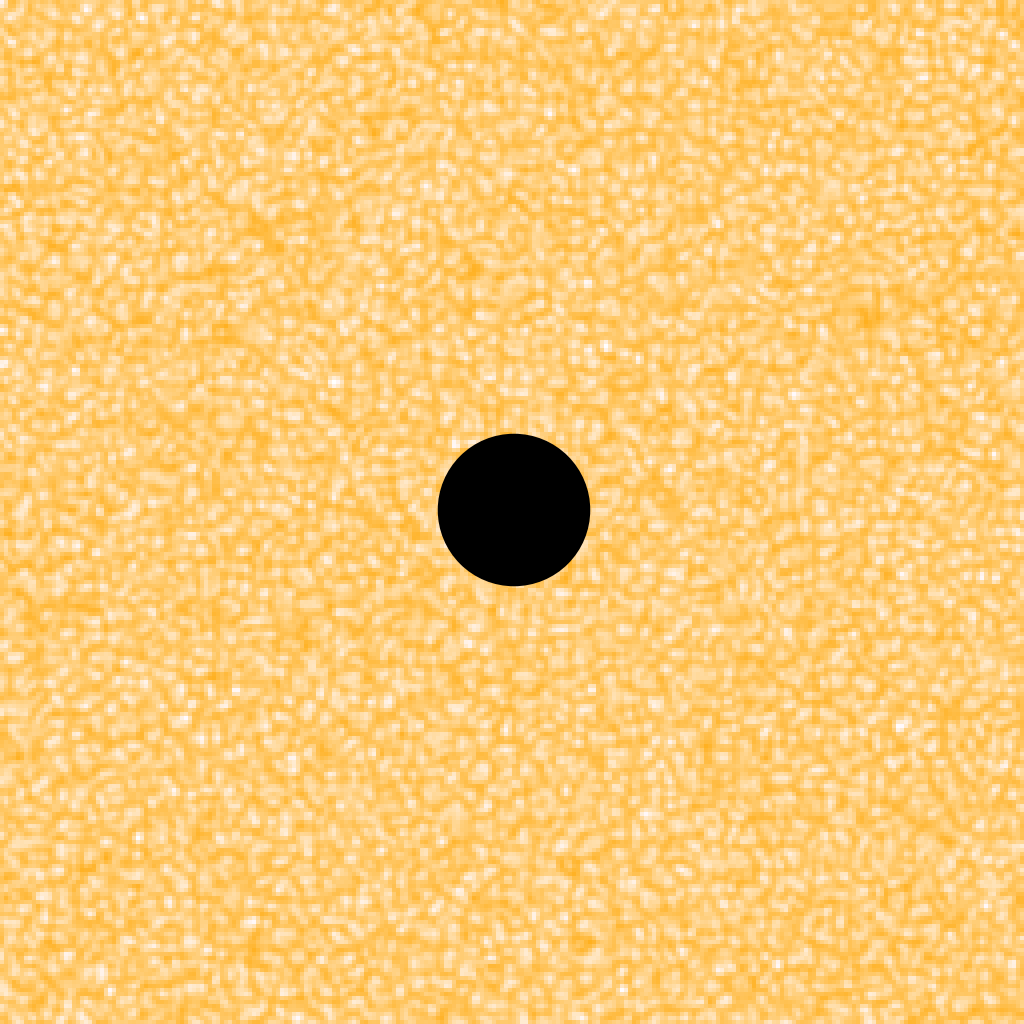}
    \caption{Same as Figure \ref{fig:hmi}, but zoomed in on a 128 x 128 arcsecond patch of the the quiet Sun at disk center. The black circle shows the scale of a simulated Earth-sized exoplanet.}
    \label{fig:hmi2}
\end{figure}

\section{Flux Variations of the Regions Occulted by the Planet} \label{sec:sdohmi}

In this section we simulate the variation in transit depth due to
spatial variations of the stellar surface brightness which is occulted
by the planet.

\subsection{Simulating transits}

\begin{figure*}
    \centering
    \includegraphics[scale=0.9]{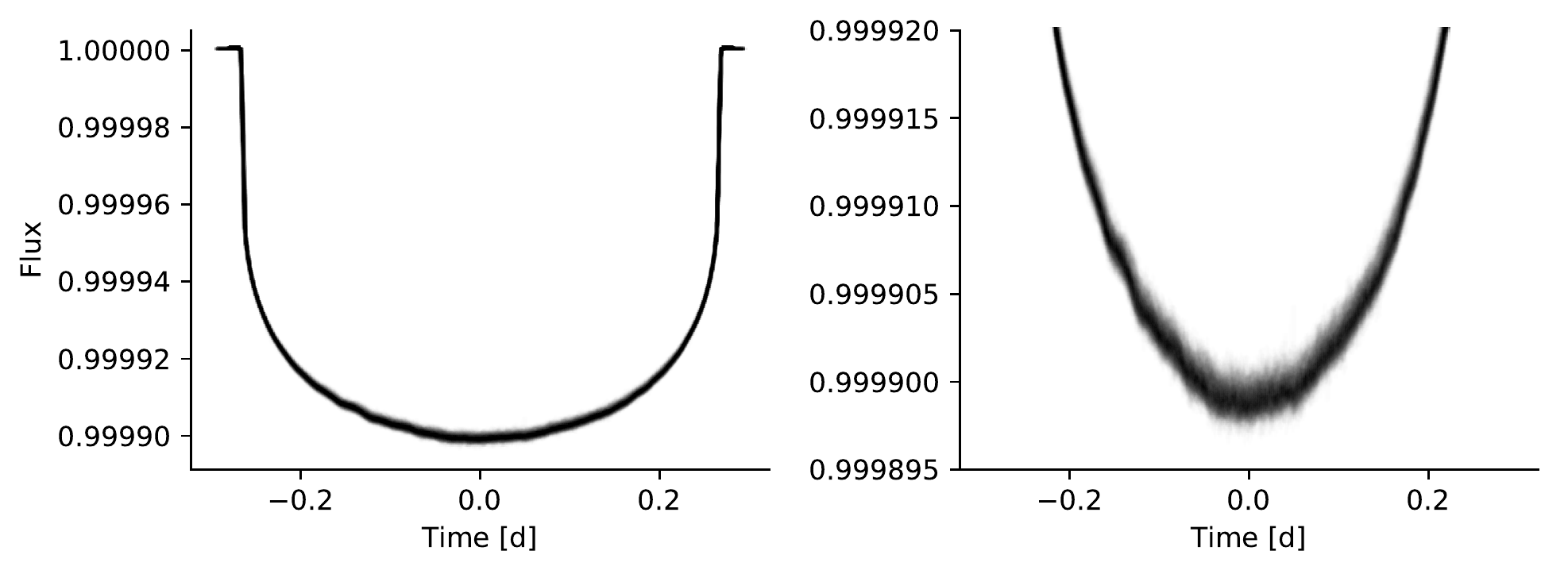}
    \caption{282 simulated transit light curves of an Earth-sized planet transiting the SDO/HMI images of the Sun (left: full transit, right: zoom into mid-transit). The spread at flux minimum is due to solar surface brightness variations due to activity and granulation.}
    \label{fig:lc}
\end{figure*}

To simulate transits, we first select a single, static continuum intensity image. We then create a time series of fluxes by first creating a synthetic exoplanet, equal to the projected size of the Earth, and summing the flux within this circular disk.  
We then subtract this sum from the total solar disk intensity. The projected size of the Earth is computed from the HMI image headers, which record the apparent solar radius, in units of pixels, for each image.  

We ignore the effects of the Moon or other planets on the orbit of Earth, and we also neglect the eccentricity of Earth's orbit for simplicity.  We advance the planet incrementally across the solar surface and compute the time of each integration assuming a circular orbit with $P_\mathrm{orb} = 365.25$ days. We create this time series across a single, static image to avoid the diurnal variation in the amplitude of the intensity measurements. This has the added benefit of eliminating the effects of photon noise (since the entire time series is across one realization of photon noise), which allows us to probe the effects of photospheric granulation at the ppm level. 

We simulate transits across 282 deconvolved continuum intensity images taken on days in 2018 that show the least magnetic activity. This ensemble of transits is shown in Figure~\ref{fig:lc}. We only simulate transits with an impact parameter of $b=0$. Selecting $b=0$ ensures that the planet is unlikely to occult magnetically active regions, which rarely occur near the solar equator, and maximizes the granulation contrast. The software for these simulations is available online\footnote{Open source software: \url{http://github.com/bmorris3/stash}}. 

\subsection{Granulation and supergranulation noise in the transit residuals} \label{sec:granulation}

\begin{figure*}
    \centering
    \includegraphics[scale=0.6]{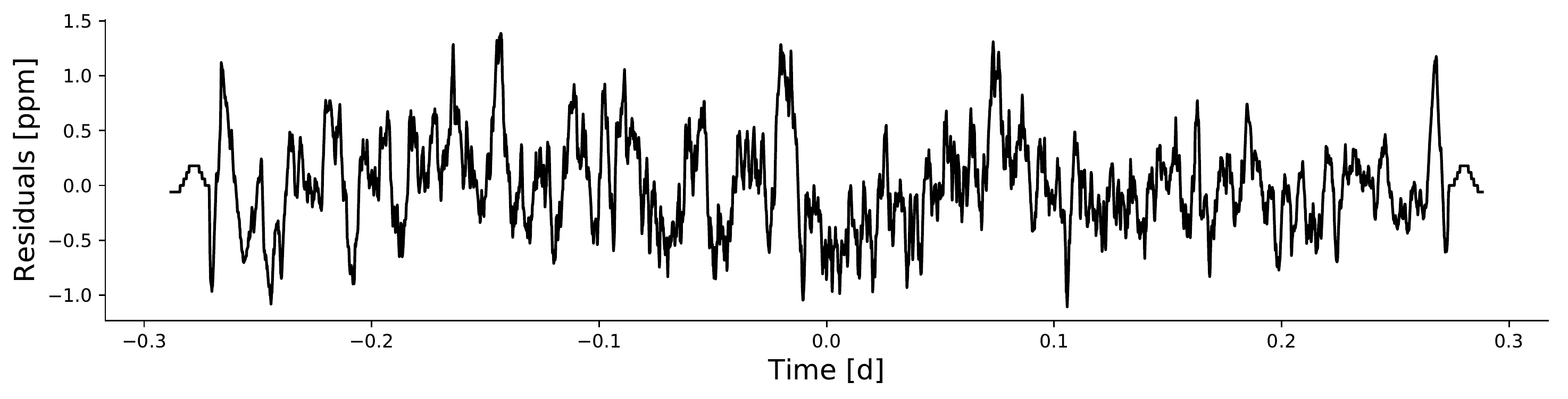}
    \caption{Transit residuals of a particular realization of the solar granulation, showing residual amplitude $\sim 2$ ppm. }
    \label{fig:example_residuals}
\end{figure*}

We fit the \citet{Mandel2002} transit model to each of the simulated light curves, implemented by the Python package \texttt{batman} \citep{Kreidberg2015}. For each light curve, we simultaneously fit for the planet radius, mid-transit time, orbital inclination and four nonlinear limb-darkening parameters, while fixing the orbital period, semi-major axis, and radius of the star at the known values, assuming a circular orbit for the Earth. We then examine the residuals of each transit fit (an example is shown in Figure~\ref{fig:example_residuals}) and reject any transits contaminated by magnetic elements (transits containing residual flux values $>5$ ppm). We used nonlinear limb-darkening parameters here to be sure to remove any symmetric trends in the light curve. To remove imperfections in the HMI deconvolved image flat field, which we discussed in Section~\ref{sec:hmi}, we subtract the residual flux at each time by the median of all transit residuals. The resulting transit residuals typically have standard deviations of 0.5 ppm, and a typical range is $2-4$ ppm.

\subsection{Radius uncertainty due to granulation and supergranulation} \label{sec:gp}

Granulation occurs on a variety of size scales -- granules with radii of $\sim 0.5$ Mm, which are clearly visible in the HMI continuum intensity images (see Figure~\ref{fig:hmi}), and supergranules, with radii of $\sim 16$ Mm. Granules and supergranules appear stochastically over time, but their length scales and turnover times stay roughly constant. Therefore, we can model our ensemble of 282 transit residuals as an autocorrelated signal with noise. 

We model the autocorrelated signal with Gaussian process regression -- for more information on Gaussian processes, see \citet{Rasmussen2006}. We use a simple harmonic oscillator kernel implemented by the Python package \texttt{celerite} \citep{Foreman-Mackey2017}, which has the following power spectrum:
\begin{equation}
S(\omega) = \sqrt{\frac{2}{\pi}} \frac{S_0 \omega_0^{4}}{(\omega^2 - \omega_0^2)^2 + \omega_0^2\omega^{2}/Q^2},
\label{eqn:sho}
\end{equation}
where $S_0$ is the amplitude, $\omega_0$ is the characteristic oscillation frequency, and $Q=1/\sqrt{2}$ is the quality factor of the oscillation.
We use a Markov Chain Monte Carlo method, implemented by the Python package \texttt{emcee} \citep{Foreman-Mackey2013}, to simultaneously fit for the simple harmonic oscillator kernel hyperparameters $S_0$ and $\omega_0$ as well as the \citet{Mandel2002} transit model parameters for the orbital inclination $i_o$, quadratic limb darkening parameters $u_1$ and $u_2$,
mid-transit time $t_0$, and $R_p/R_\star$, the ratio of the exoplanet radius, $R_p$, to the stellar radius, $R_\star$.

We find a typical uncertainty on the exoplanet radius is 0.02\% $R_p$. We consider this a noise floor, or a lower limit, since we modelled the granulation pattern from a synthetic transit across a single, static image. 

We also find that the transit residuals are roughly periodic with timescale $P = 18 \pm 1$ minutes. We can use this periodicity to measure the characteristic length scale for supergranulation. Assuming a supergranule is large compared to the planet, the duration of a supergranule occultation $\tau$ by a small transiting exoplanet (with an impact parameter $b=0$) is approximately

\begin{equation}
\tau \approx \frac{2R_\mathrm{sg}}{v} = \frac{R_\mathrm{sg} P_\textrm{orb}}{\pi a} \label{eqn:tau}
\end{equation}

where $R_\mathrm{sg}$ is the radius of the supergranule, $v$ is the orbital velocity of the planet, $P_\textrm{orb}$ is the orbital period of the planet, and $a$ is the semimajor axis of the planet's orbit. Rearranging Equation~\ref{eqn:tau} for the supergranule radius, we find $R_\mathrm{sg} = 16 \pm 1$ Mm, similar in horizontal scale to supergranules observed both in radial velocity maps of the solar surface \citep{Hathaway2000} as well as in continuum intensity \citep[see for example][]{Meunier2008, Goldbaum2009, Rieutord2010}.  Modelling the granulation across a time-varying background would account for the convective motions of the granules and therefore introduce a more complex pattern for the granulation noise. However, supergranulation imprints itself in patterns on the Sun with characteristic lifetimes of $\sim 1.8$ days \citep{Rieutord2010}, which is longer than a typical Earth transit of a Sun-like star ($\sim 12$ hours), and so the static approximation should be adequate.

\subsubsection{Comparison with previous results}

\citet{Chiavassa2017} simulated transits of Earth-like planets on three-dimensional radiative hydrodynamical simulations from the {\sc Stagger} grid models of Sun-like stars and found the residual signal in the transit due to granulation had RMS amplitude 3.5 ppm in the bandpass 7600-7700 \AA, which is similar to our estimate using HMI continuum intensity images ($2-4$ ppm).

\subsubsection{Expectations for other stars}

Numerical simulations of stellar granulation for stars across the main sequence show that granule size scales inversely with the stellar surface gravity \citep[see review by][]{Nordlund2009, Kupka2017}. As a result, one might expect that stars smaller than the Sun will have smaller granules. The smaller the granules, the more granules occulted by an Earth-sized planet in a given exposure, and therefore the smaller the in-transit signal of granulation on the light curve. 

Numerical simulations of granulation for stars from spectral type F7-K3 dwarfs by \citet{Trampedach2013} all have characteristic horizontal scales of granules of order 1 Mm. These scales grow as stars evolve and their $\log g$ decreases \citep[see also][]{Beeck2013b, Beeck2013a, Trampedach2017}. Therefore the small-scale granulation signal should be most significant for evolved stars.

Solar supergranulation, in contrast with small scale granulation, is difficult to measure due to its small amplitude, and difficult to simulate due to its vast physical extent \citep{Rieutord2010}. Thus we caution the reader to only use the results of this analysis for solar twins.

\section{Flux Variations of the Unocculted Stellar Disk} 
\label{sec:photometry}

\begin{figure*}
    \centering
    \includegraphics[scale=0.6]{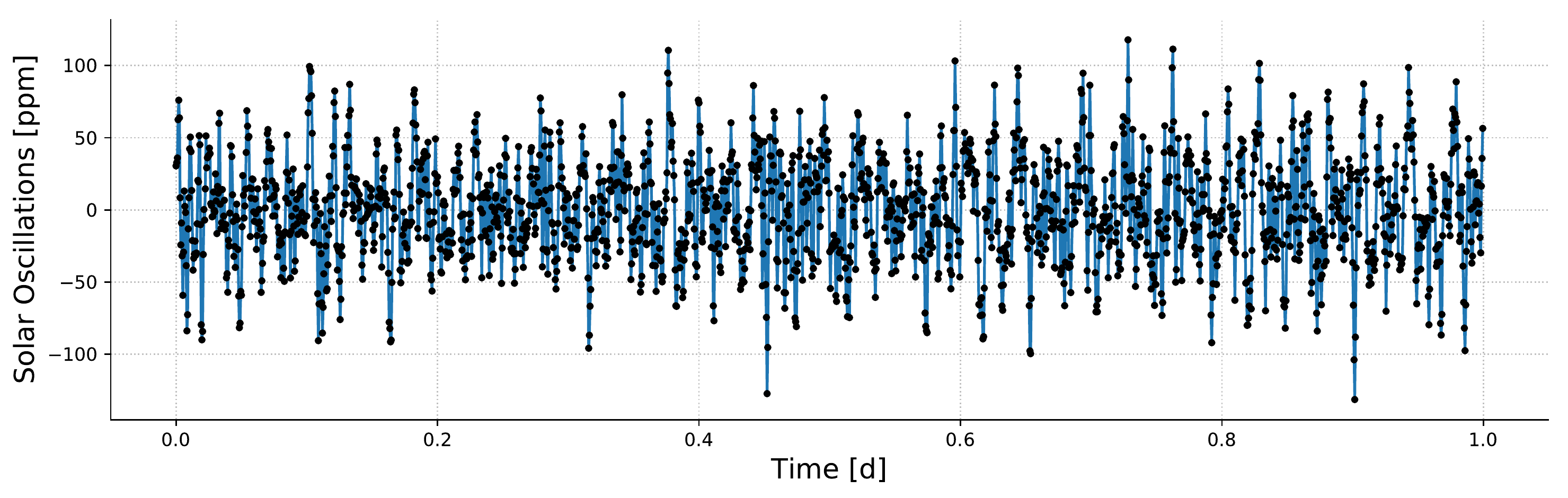}
    \caption{Twenty-four hours of 45-second cadence photometry of the Sun from SDO/HMI continuum intensity images, with the mean intensity removed to show the small amplitude variability of the Sun on these timescales (black circles). The error bars on the fluxes from each measurement are smaller than the points. The blue curve is the maxmimum-likelihood Gaussian process fit to the light curve with a simple harmonic oscillator kernel. }
    \label{fig:pmodephot}
\end{figure*}

In Section~\ref{sec:sdohmi} we studied the effect of stellar brightness variations due to granulation within the transit chord of an Earth-like exoplanet on the transit light curve. We assumed that the disk-integrated brightness of the star was unchanging, and only the spatial surface brightness variations within the transit chord were responsible for perturbations to the transit residuals. While this was a pedagogically interesting exercise, it was fundamentally one step removed from real photometry of stars: $p$-mode oscillations, magnetic activity and granulation inject time variability into the disk-integrated flux of a Sun-like star. In this Section, we measure the amplitude of the variability of the unocculted star.  Given that the transit depth is of order 0.01\%, the planet only occults a small fraction of the stellar disk, and so this variability will be present at the same level both in transit and out of transit.

\subsection{Simulating transits}

First, we measure the total brightness of the Sun using 45-second cadence HMI continuum intensity images throughout four sets of 24-hour observations, each separated by one year. The four sets of observations represent one transit observation for each of the four years of the nominal PLATO mission. This disk-integrated variability will affect both in- and out-of-transit photometry observed by PLATO. We compute the total intensity of each HMI image taken on 21 January 2018, a day with little magnetic activity near solar minimum, as well as the same day in 2015, 2016, and 2017. 

The resulting photometry has a strong diurnal signal, due to the orbital velocity of the SDO spacecraft (see Section 2.10 of \citealp{couvidat16}), which dominates over the small-scale perturbations we are trying to measure in this work. Our second step is to remove this trend by modelling the light curve with a smooth Gaussian process using a Mat{\'e}rn 3/2 kernel \citep{Rasmussen2006}, and divide the light curve by the maximum-likelihood fit, which had $\sigma= 0.00409$ and $\rho = 15$ ${\rm hours}$.

The resultant systematics-corrected HMI photometry for 21 January 2018 is shown in Figure~\ref{fig:pmodephot} (black circles), with the mean flux from the time series removed. The amplitude of the variability in the 45-second cadence photometry is $\sim$100 ppm. We also show the maximum-likelihood Gaussian process fit with a simple harmonic oscillator kernel (blue curve). 

\begin{figure*}
    \centering
    \includegraphics[scale=0.75]{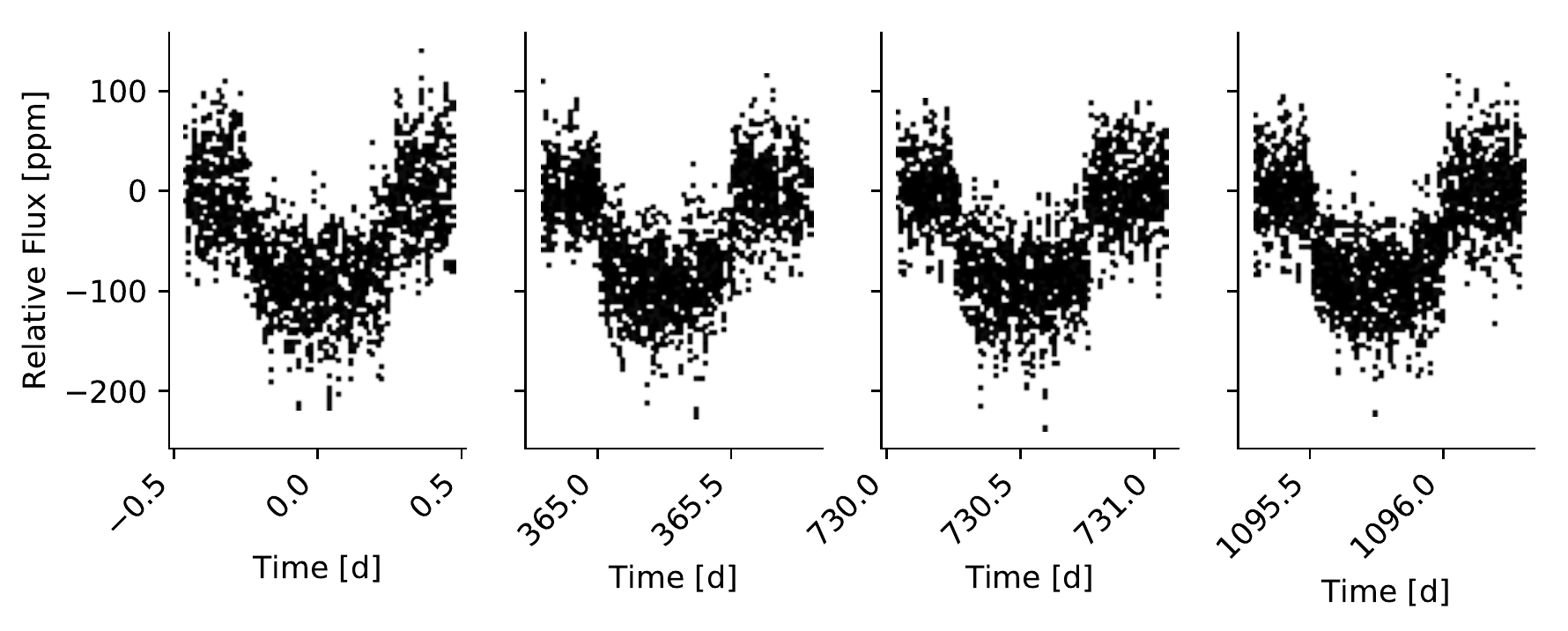}
    \caption{Transits of an Earth-sized planet across the solar surface, including the disk-integrated variability due to $p$-mode oscillations and granulation. Clearly the disk-integrated variability is greater than the in-transit granulation signal discussed in Section~\ref{sec:sdohmi} (1 ppm).}
    \label{fig:transitphotometry}
\end{figure*}

Next we inject exoplanet transits with Earth's size and orbit (with $e=0$), shown in Figure~\ref{fig:transitphotometry}, into the four 24-hour sets of systematics-corrected photometric time series measurements. To do this, we could superimpose a synthetic planet on each image, as we did for a single, static image in Section~\ref{sec:sdohmi}, compute the total intensity per image, and remove the diurnal signal by modelling the light curve with a smooth Gaussian process using a Mat{\'e}rn 3/2 kernel. However, this approach may remove the small-scale perturbations we are trying to measure. Instead, we use a model to construct a synthetic \citep{Mandel2002} transit model and we multiply this with each of the four 24-hour sets of systematics-corrected photometric time series measurements as observed by HMI. 

%The model neglects the effects of foreshortening on the $p$-mode oscillations; however, since the radial component of $p$-mode oscillations is much greater than its transverse component, the strongest $p$-mode oscillation signals occur near disk center.

These simulated transits indicate what PLATO might observe if it discovers a true Earth-analog orbiting a Sun-like star. The scatter in the plots in Fig.\ \ref{fig:transitphotometry} is due to solar oscillations and granulation. The amplitude of the oscillations is similar to the transit depth of the Earth. In the HMI photometry, the error bars are similar in scale to the size of the points in the plot.  We have not added additional noise to these
simulations, and so the photometric variability is the fundamental limit on the precision of the transit model.
 
\subsection{Radius uncertainty} \label{sec:gp2}

\begin{figure}
    \centering
    \includegraphics[scale=0.8]{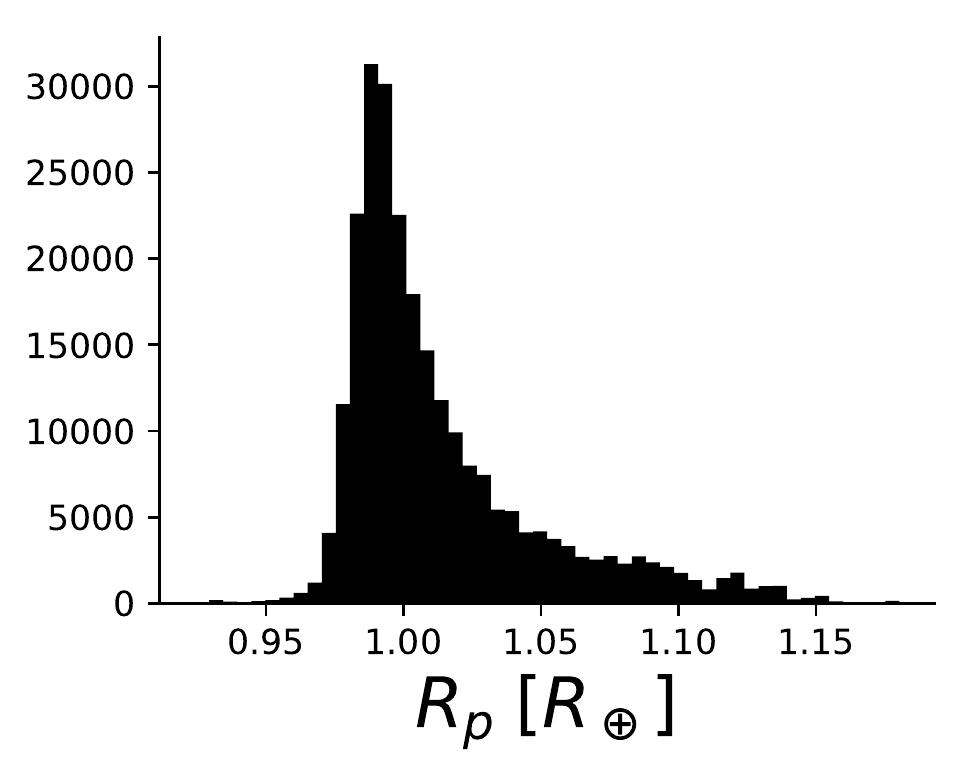}
    \caption{Posterior distribution of the radius of the Earth-sized planet, in units of Earth radii. The uncertainty is 3.6\%, accounting for the full noise budget of the PLATO mission, before any instrumental or systematic effects are accounted for. The dominant  contributor towards this uncertainty is degeneracy with impact parameter in the presence of noise due to stellar oscillations and granulation (see Appendix~\ref{sec:posteriors} for full posterior distributions).}
    \label{fig:radius_posterior}
\end{figure}

\begin{figure}
    \centering
    \includegraphics[scale=0.7]{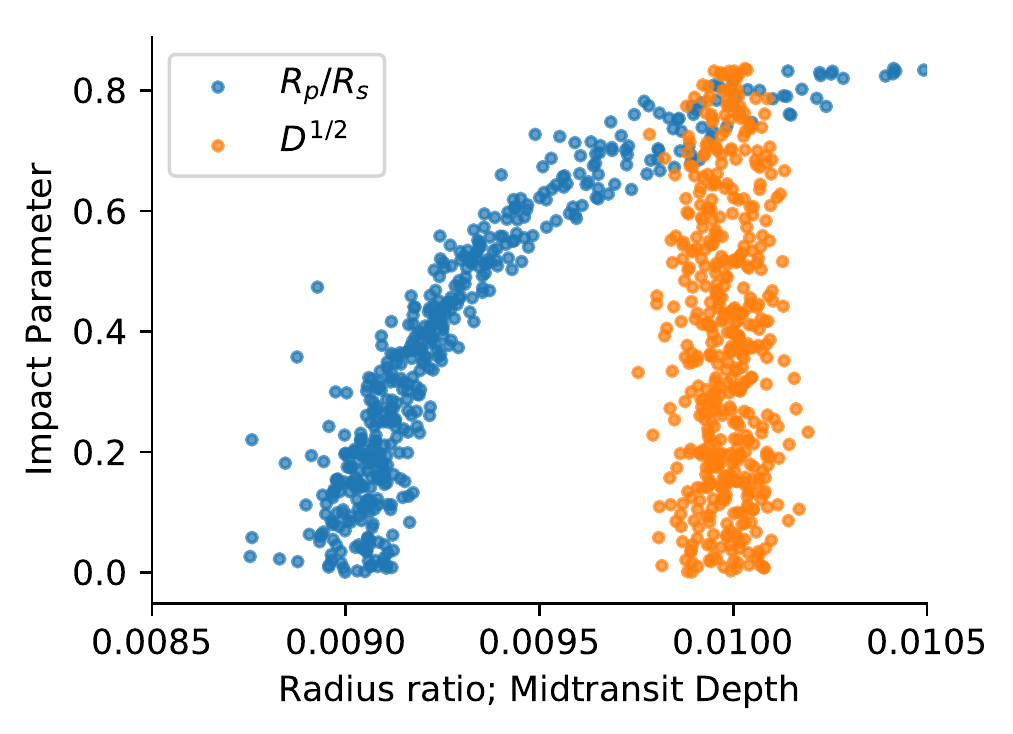}
    \caption{Posterior distribution of radius-ratio, $R_p/R_*$, versus impact parameter (blue) and the square root of depth, $D^{1/2}$ versus impact parameter (orange).  The uncertainty on $D^{1/2}$ is 0.73\%}
    \label{fig:radius_depth_impact_parameter}
\end{figure}

To determine the uncertainty in an exoplanet's radius due to solar oscillations and granulation, we follow the same procedure as in Section~\ref{sec:gp}. We simultaneously fit a \citet{Mandel2002} transit light curve and a Gaussian process with a simple harmonic oscillator kernel to the transit photometry in Figure~\ref{fig:transitphotometry} to measure the uncertainty in the exoplanet radius while accounting for the correlated noise due to oscillations and granulation. In the previous section, the periodicity measured by the kernel arose from the synthetic planet spatially crossing bright convective upflows and dark inter-granular lanes. In this case, the periodicity measured by the kernel is due to $p$-mode oscillations and granulation. In this section, we also let the orbital period float, so as to let its uncertainty propagate into the uncertainties on other parameters since the period will be unknown for newly discovered transiting planets.

The posterior PDF for the planet radius is shown in Figure~\ref{fig:radius_posterior}, for
which the uncertainty in the planet radius is 3.6\%. This uncertainty is the combined effect of three contributions: the degeneracy with the unknown impact parameter, and the signals of stellar granulation and oscillations.  The square-root of the maximum depth of transit, $D^{1/2}$, is measured with much higher precision, about 0.73\%, as shown in
Figure~\ref{fig:radius_depth_impact_parameter}, where depth is defined as:
\begin{equation}
    D = \frac{R_p^2}{R_*^2} \frac{1-u_1(1-\sqrt{1-b^2})-u_2 (1-\sqrt{1-b^2})^2}{1-\tfrac{u_1}{3}-\tfrac{u_2}{6}},
\end{equation}
where $R_p$ is the planet radius, $R_*$ is the stellar radius, $u_1$ is the linear limb-darkening parameter, $u_2$
is the quadratic limb-darkening parameter, and $b$ is the impact parameter mid-transit (in units of the stellar
radius).  The significant limb-darkening of the star in the PLATO passband, which we have approximated with
the limb-darkening coefficients in the \kepler band given by $u_1=0.45$ and $u_2=0.19$ \citep{Sing2010}, is the origin of the
degeneracy between impact parameter and radius ratio, which is responsible for the long tail in radius-ratio
seen in Figure \ref{fig:radius_posterior}.  The simulation was created with an edge-on orbit, corresponding
to $b=0$, while in the presence of stellar noise, the recovered impact parameter varies from 0 to 0.8 (Figure~\ref{fig:radius_depth_impact_parameter}).
At larger values of the impact parameter, the duration of the transit shortens and the depth of transit
becomes shallower due to the lower surface brightness of the star.  However, the {\it shape} of
the trough of the transit is very similar shape as zero impact parameter.  Thus, at higher impact
parameter, very good fits to the data are obtained by adjusting the duration of the transit 
(this comes from adjusting the ratio $a/R_*$), and by increasing the radius ratio of the planet to the
star to maintain the same transit depth, as well as a very slight change in the limb-darkening coefficients.  

We have also performed fits over a broad range of impact parameters from $b=0 - 0.9$, and find that the results remain unchanged. The constraints on impact parameter are weakest near $b=0$, where the limb-darkening coefficients matter less, and the constraints on the limb-darkening coefficients are greatest when $b\rightarrow 0.9$, where the impact parameter is better constrained by the light curve. As a result, the radius precision does not change significantly with $b$.

The one quantity that changes significantly at higher impact parameters
is the duration of ingress and egress, which is smaller than the transit duration by a factor of
$(R_p/R_*)(1-b^2)^{-1}$.  Although
the higher impact parameters have a longer duration of ingress and egress, the uncertainty on the measurement of
this duration from the data is significant due to the
noise caused by stellar oscillations and granulation variability, which have a similar characteristic timescale.
It is only at very large impact parameter, $b \approx 0.8$, that the ingress/egress duration becomes sufficiently long that the
data can rule out the model with a large impact parameter; this is why the posterior distribution cuts off
at high impact parameter.  Unfortunately there is a significant variation in the radius ratio
as a function of impact parameter, and so the lack of constraint upon the impact parameter
yields a much larger uncertainty on the radius ratio compared with the uncertainty on the square root of transit depth, $D^{1/2}$.

To further isolate the contributions of the uncertainty in impact parameter/inclination, we fixed the impact parameter to $b=0$ and refit the light curve.  This yielded a radius uncertainty of 0.5\%, which is a {\it much} smaller uncertainty than in the
variable inclination fit.  This indicates that the majority of the measured uncertainty in the radius ratio
actually arises from degeneracy with the impact parameter in the presence of noise due to granulation and oscillations.

The degeneracy between the planet radius and the unknown impact parameter may present a significant obstacle to precision exoplanet radius characterization with PLATO, but it still may be possible to mitigate the impact of this degeneracy. As with the procedure presented in  \citet{Morris2018}, asteroseismic analysis of the light curve can yield precise estimates of the mean stellar density, which constraints $a/R_\star$ via Kepler's third law. The measured duration of the transit in turn constrains the impact parameter via
\begin{equation}
    b = a/R_\star \cos(i) \left(\frac{1-e^2}{1+e\sin{\omega}}\right)
\end{equation}
\citep{Winn2010b}, with some uncertainty due to the unknown eccentricity of the orbit, $e$, and longitude of pericenter, $\omega$. Therefore by measuring the asteroseismic signal of the star, as PLATO is designed to do, better precision on the impact parameter, and thus the planet radius, could be achieved with a prior on the eccentricity.  Another means of characterizing the stellar density would be to use the transits of other planets in the system \citep{Kipping2014}, which would similarly constrain the impact parameter.  If the eccentricity of the orbit could be measured through other means, such as radial-velocity monitoring of the star, then the constraint upon the impact parameter could be tightened, and the full precision of the radius ratio recovered. Another complementary measurement which in principle could help constrain the impact parameter is the Rossiter-McLaughlin effect, though this will be difficult to measure in practice for Earth-sized planets.

The limb-darkening parameters contribute to the radius uncertainty, as the limb-darkening in a broad optical bandpass like PLATO's will be significant.  Our fits to simulated photometry show that the posterior PDFs of the limb-darkening parameters are somewhat constrained (see Appendix~\ref{sec:posteriors} for full posterior distributions), and are not strongly covariant with the impact parameter, and so the degeneracy between impact parameter (or inclination) and limb-darkening remains.  If follow-up observations
were obtained in the infrared, where limb-darkening is much less significant, then the exoplanet radius can be measured with less dependence upon the impact parameter.  An additional advantage of observing in the infrared would be quieter variability of the star,
and a stronger variation of the transit light curve at ingress and egress, which may yield a further constraint upon the impact parameter based upon the shape of the light curve.

The predicted uncertainty, due to degeneracy between impact parameter and exoplanet radius in the presence of stellar granulation and oscillations, amounts to the full error budget in the mission specifications for PLATO. Marginal improvements on the precision of the transit light curves may be obtained by modeling more of the out-of-transit light curve, or with an extended mission which observes the same field for more transits. 

% The oscillation period of the maximum-likelihood Gaussian process is $P=5.83 \pm 0.10$ minutes. This is slightly longer than the canonical ``five-minute oscillation'' period of $5.39 \pm 0.05$  min \citep{Fröhlich1997,Huber2011}.

\section{Discussion} \label{sec:discussion}

The PLATO mission seeks to measure Earth-analog exoplanet radii with 3\% precision. We have demonstrated, using simulated photometry from HMI continuum intensity images, that stellar oscillations and granulation coupled to degeneracy between the planet radius and orbital impact parameter caused by limb-darkening will be important contributors to the uncertainty in exoplanet radii, of order 3\%. In order to reach 3.6\% precision, we took advantage of the extremely high signal-to-noise of the HMI observations and fit a Gaussian process to the solar oscillations and granulation.

The HMI photometry detrending strategy in Section~\ref{sec:photometry} used to remove the continuum intensity trends with orbital phase likely also removed solar variability on timescales greater than a few hours. As a result, there may be additional sources of correlated noise that are not incorporated into the radius uncertainties we report in this work. Primarily, we note that super- and meso-granulation are sources of correlated noise on the hours-to-days timescales similar to the transit of an Earth across a Sun \citep{Fröhlich1997, Aigrain2004a}. As such, the 3.6\% radius uncertainty noise floor that we present here should be understood as a lower limit -- incorporating the longer timescale sources of variability will increase the radius uncertainty. 

We carried out a test of whether the detrending technique is removing significant signals that will affect the exoplanet radius uncertainty.  We downloaded four complete days of continuous, 1-minute cadence continuum intensity images taken by the Michelson Doppler Imager (MDI; \citealt{scherrer95}) aboard the Solar and Heliospheric Observatory (SoHO). MDI took data from 1996 until 2010 from the L1 Lagrange point, and thus its continuum intensity data does not show the orbital phase variations present in the SDO data. Therefore, the MDI observations\footnote{MDI observations are publicly available at \url{http://jsoc.stanford.edu}} are an ideal control data set for comparison against the detrended HMI continuum intensity photometry. We find that after injecting transits into the MDI observations, the posterior distributions for the exoplanet radius have a similar uncertainty as the detrended HMI observations (4\%). This indicates we do not significantly underestimate the uncertainty in the exoplanet radius due to the HMI detrending process.

In our analysis of the Earth-like transit light curves, we assumed the period was known perfectly $P_{orb}=365.25$ d, and that no transit timing variations occurred. In real observations of the Earth, the uncertain period and transit timing variations would introduce larger uncertainties on the impact parameter, and therefore the radius ratio.

\subsection{Wavelength dependence of variability}

In the calculations presented above, we assumed that the variability that PLATO will observe is similar to the ``psuedo-continuum'' variability observed by SDO/HMI. However, HMI continuum intensity is observed over a narrow $\sim 1$ \AA\ bandpass centered on 617.3 nm, whereas PLATO will observe in a broad bandpass from 500-1050 nm. Therefore we must verify that the variability observed in the narrow HMI bandpass is not significantly different from the variability extrapolated into the PLATO bandpass. We can approximate the ratio of variability in the HMI bandpass to the variability in the PLATO bandpass as follows.

Ignoring limb-darkening, let's say a fraction $f_c \ll 1$ of the star is covered by inter-granular lanes with a temperature $T_c$, while the remainder of the star we take as having a constant temperature, $T_s$.
Then, the flux from the star is given by:
\begin{equation}
F _ { \nu } = \Omega \left[ f _ { c } I _ { \nu } \left( T _ { c } \right) + \left( 1- f _ { c } \right) I _ { \nu } \left( T _ { s } \right) \right]
\end{equation}
where the flux can vary in the \kepler band is via variations in either $T_c$ or $f_c$.  The specific intensity is $I_\nu(T)$ for an atmosphere at temperature $T$, while $\Omega$ is the solid angle of the star.   We Taylor expand the intensity in terms of temperature:
\begin{equation}
    I_{\nu}(T_C) = I_\nu(T_S) + (T_C-T_S) \frac{\partial I_\nu}{\partial T}\bigg\vert_{T_S},
\end{equation}
giving
\begin{equation}
F _ { \nu } \approx \Omega \left[I _ { \nu } \left( T _ { s } \right)+ f _ { c } (T_C-T_S) \frac{\partial I_\nu}{\partial T}\bigg\vert_{T_S}   \right]
\end{equation}

Then, integrating over the SDO/HMI (subscript $H$) and PLATO (subscript $P$) bands:
\begin{eqnarray}
\dot { N } _ { H } &=& \int d \nu \frac { F _ { \nu } } { h \nu } T _ { H ,\nu } \\ 
\dot { N } _ { P } &=& \int d \nu \frac { F _ { \nu } } { h \nu } T _ { P ,\nu }
\end{eqnarray}
where $T_{H,\nu}$ is the SDO/HMI throughput at frequency $\nu$, and $\dot{N}_H$ is the photon count rate (cm$^2$ s$^{-1}$).

When $F_\nu$ varies, this causes variation in $\dot{N}_H$ and $\dot{N}_P$, but both of these just depend on $f_c(T_C-T_S)$. Since the amplitude of variation is small, we approximate
\begin{multline}
\frac { d \dot { N } _ { H } } { d f _ { c } } \frac { 1} { \dot { N } _ { H } | _ { f _ { c } ,0} } \approx \left( T _ { c } - T _ { s } \right) \frac{ \int d \nu \frac{\partial I _ { \nu }}{\partial T}\big\vert_{T_S}  \nu ^ { - 1} T _ { H ,\nu } } { \int d \nu I _ { \nu } \left( T _ { s } \right) \nu ^ { - 1} T _ { H ,\nu } }
\end{multline}
where the approximation assumes $f_c \ll 1$ and $T_C-T_S \ll T_S$. A similar relation holds for the PLATO bandpass. Thus, the ratio of the fractional amplitude of
variation in PLATO to that in SDO/HMI, $\alpha$, is given by:
\begin{multline}
\alpha \equiv \left( \frac { d \dot{N} } { d f _ { c } } \frac { 1} { \dot { N }_P | f _ { c ,0} } 
\right)  \left( \frac { d \dot{N} _ { H } } { d f _ { c } } \frac {1} { \dot{N} _ { H } | f _ { c ,0} } \right)^{-1} \approx\\
\frac { \int d \nu \frac{\partial I _ { \nu }}{\partial T}\big\vert_{T_S} \nu ^ { - 1} T _ { P ,\nu } } { \int d \nu \frac{\partial I _ { \nu }}{\partial T}\big\vert_{T_S} \nu ^ { - 1} T _ { H ,\nu } } \frac { \int d \nu I _ { \nu } \left( T _ { s } \right) \nu ^ { - 1} T _ { H ,\nu } } { \int d \nu I _ { \nu } \left( T _ { s } \right) \nu ^ { - 1} T_{P ,\nu}},
\end{multline}
where, conveniently, the dependence upon $f_c$ and $T_C$ has disappeared.

Approximating the Solar spectrum as a blackbody, 
we estimate that the variability of a Sun-like star with $T_\mathrm{eff} = 5777$ K as observed in the PLATO bandpass will have 86\% of the variability observed in the SDO/HMI bandpass. We therefore assume that the variability due to granulation estimated by SDO/HMI is a fair proxy for the variability that will be observed by PLATO, or slightly overestimated.

\subsection{Limb-darkening parameters}

Given that the design of the PLATO instrument has yet to be finalized at the time of writing, we have used the limb-darkening parameters computed for the {\it Kepler} bandpass as a stand-in for the limb-darkening in the PLATO band.  We justify this by estimating the expected quadratic limb-darkening parameters in
the PLATO band.

We utilize the radiation hydrodynamic ``stagger-grid" models of \citet{Magic2015}
to estimate the limb-darkening in the PLATO bandpass.  We assume
that the PLATO transmission is constant over the 500-1050 nm,
and we integrate the photon-weighted specific intensity from the averaged models
of \citet{Magic2015}\footnote{Downloaded as an IDL .sav file, \texttt{mmu\_t5777g44m00v05.flx}, from \texttt{https://staggergrid.wordpress.com/clv/}.  }.  Note that
this model has been calibrated to the Sun, and gives limb-darkening
which closely approximates the Sun \citep{Magic2015}.

We find best-fit quadratic limb-darkening parameters for the Solar model over this
wavelength range of $u_1 = 0.24$ and $u_2 = 0.36$.  This compares favorably
with the Kepler limb-darkening parameters which we have used in our
simulations;  the stagger-grid surface brightness only differs from the
quadratic Kepler surface brightness by a few percent over the range of
observed inclination angles.

\section{Conclusion} \label{sec:conclusion}

We have constructed photometry of the Sun using HMI continuum intensity time series images to study the effects of solar $p$-mode oscillations and granulation on transit photometry. The short timescale ($<1$ hour) scatter in a transit light curve of an Earth-like planet transiting a Sun-like star has two components. First, both the in- and out-of-transit residual scatter show disk-integrated temporal brightness fluctuations driven by $p$-mode oscillations and granulation, of order 100 ppm in amplitude with a five-minute period. 

Second, maps of stellar surface brightness variations within the transit chords of exoplanets are encoded in the residuals of transit photometry. We find that the brightness variations due to granulation impart only a slight additional scatter, of order 2-4 ppm, to the in-transit photometry, in good agreement with studies of simulated stellar atmospheres by \citet{Chiavassa2017}.

We demonstrate that transiting exoplanet radius uncertainties of 3.6\% are possible with photon noise-limited photometry from HMI continuum intensity images by accounting for the correlated disk-integrated $p$-mode and granulation signals with a Gaussian process. This uncertainty on Earth-like exoplanet radii due largely to degeneracy with impact parameter accounts for the full error budget for the PLATO mission, and perhaps motivates: (1) careful propagation of the stellar density measured via asteroseismology into the measurement of the exoplanet's orbital parameters; (2) auxiliary observations of PLATO targets in the infrared where stellar variability and limb-darkening are weaker to measure precision exoplanet radii and orbital parameters; and (3) a longer extended mission to improve upon the radius measurement uncertainties. We emphasize that these simulated observations focus on observations near solar minimum, and therefore represent conservative estimates on the noise contributed by the Sun and Sun-like stars. This kind of analysis may also help characterize fundamental noise floors for transits observed by the Transiting Exoplanet Survey Satellite (TESS) and CHaracterizing ExOPlanet Satellite (CHEOPS) missions. 

In general, the radius-impact parameter degeneracy discussed in Section~\ref{sec:gp2} should affect smaller planets more strongly than large planets. The duration of ingress and egress in the limit of small impact parameter scales linearly with the radius ratio $R_p/R_\star$, so the shape of the light curve, and thus the constraint on the planet radius, is weaker for smaller planets.  Thus, we expect that our results will apply more generally for observations of small transits at wavelengths with strong limb-darkening.

\section*{Acknowledgements}

We are grateful to Manodeep Sinha, who helped make the numerical methods presented in this work vastly more efficient. We would also like to thank Jeneen Sommers for generating the deconvolved HMI continuum intensity data and Charles Baldner for discussions about the HMI instrument. This research has made use of NASA's Astrophysics Data System.
We gratefully acknowledge the following software packages which made this work possible: \texttt{astropy} \citep{Astropy2013, Astropy2018}, \texttt{ipython} \citep{ipython}, \texttt{numpy} \citep{VanDerWalt2011}, \texttt{scipy} \citep{scipy},  \texttt{matplotlib} \citep{matplotlib}, \texttt{celerite} \citep{Foreman-Mackey2017}, \texttt{sunpy} \citep{sunpy}, \texttt{emcee} \citep{Foreman-Mackey2013}, \texttt{corner} \citep{Foreman-Mackey2016}. 

\bibliographystyle{mnras}
%\bibliography{bibliography}

\appendix

\section{Typical posterior distributions} \label{sec:posteriors}

\begin{figure*}
    \centering
    \includegraphics[scale=0.3]{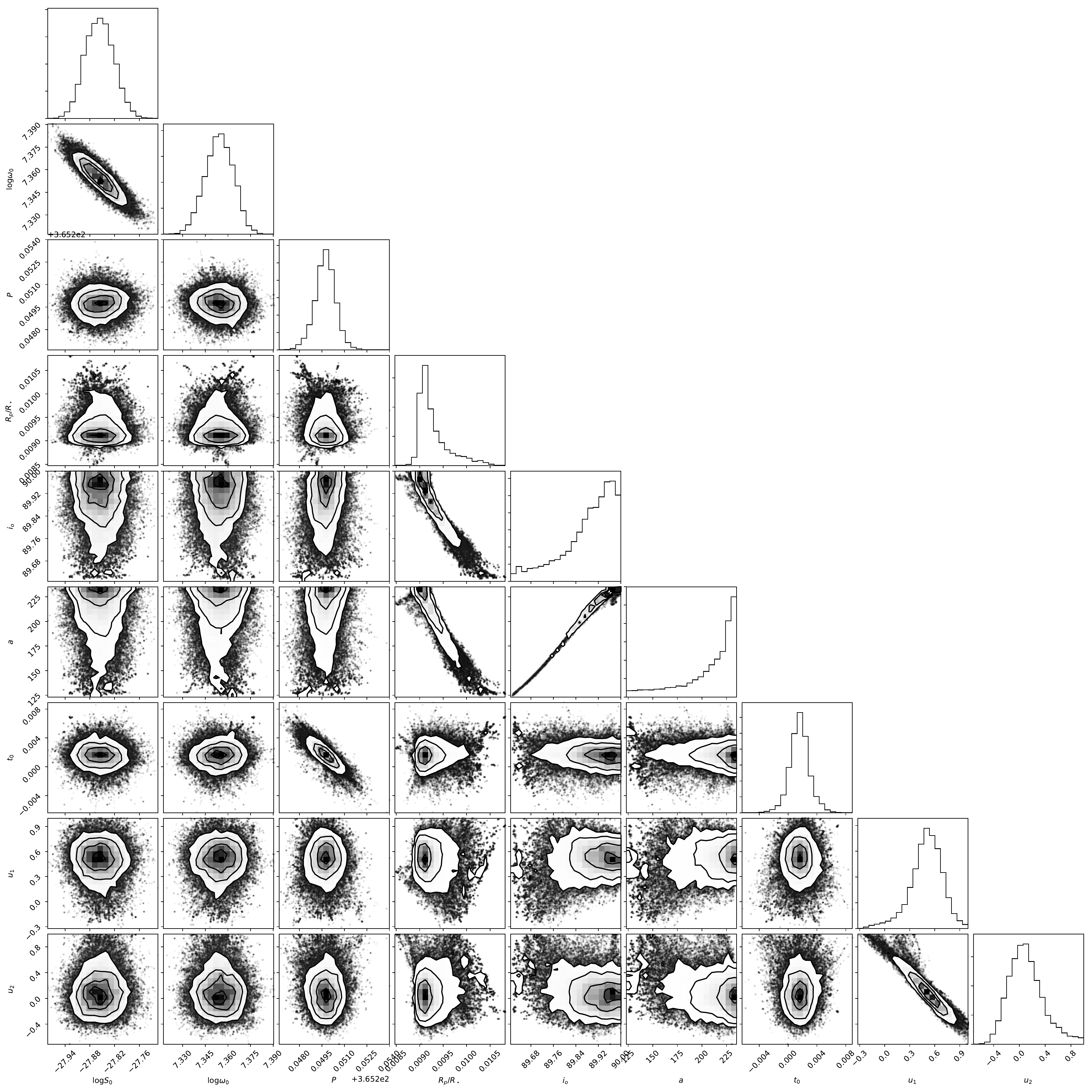}
    \caption{Posterior distributions for the Gaussian process hyperparameters and the \citet{Mandel2002} transit model parameters in a typical fit to a simulated transit. The parameters are: the GP hyperparameters for the amplitude $\log S_0$ and oscillation frequency $\log \omega_0$, the orbital period $P$, the radius ratio $R_p/R_\star$, the orbital inclination $i_o$ [degrees], the semimajor axis in units of the stellar radius $a/R_\star$, the mid-transit time $t_0$ [days], and the quadratic limb-darkening parameters $u_1$ and $u_2$. The periodicity in the residuals has timescale $P = 5.83 \pm 0.10$ minutes, and the uncertainty in the planet radius is 3.6\%. Note the strong degeneracy between $R_p/R_\star$ and $a$ and $i_o$. Convergence was assessed with the autocorrelation length of the chains.}
    \label{fig:corner}
\end{figure*}

In Section~\ref{sec:gp2} we injected a \citet{Mandel2002} transit model with the properties of Earth (impact parameter $b=0$) into the SDO/HMI continuum intensity photometry of the Sun, whose scatter is dominated by $p$-mode oscillations (see Figure~\ref{fig:transitphotometry}). We fit simultaneously for the transit light curve parameters $R_p/R_\star$, $i_o$, $a$, $t_0$, $u_1$ and $u_2$, in addition to the simple harmonic oscillator kernel hyperparameters $S_0$ and $\omega_0$, fixing $Q=1/\sqrt{2}$. In Figure~\ref{fig:corner} we show the posterior PDFs for each of the parameters. 

% Don't change these lines
\bsp	% typesetting comment
\label{lastpage}
\end{document}